\documentclass[11pt]{article}
\usepackage{amsmath}
\usepackage{amsfonts}

\usepackage{ulem}

\newcommand{\comm}[1]{}

\newcommand{\bx}{{\bf x}}
\newcommand{\bv}{{\bf v}}
\newcommand{\bp}{{\bf p}}
\newcommand{\bA}{{\bf A}}
\newcommand{\bB}{{\bf B}}
\newcommand{\pd}{{\partial}}
\newcommand{\ri}{{\rm i}}
\newcommand{\hth}{{\hat{h}}}
\newcommand{\fhh}{\mathfrak{\hat{h}}}

\begin{document}

\begin{titlepage}
\vskip 2cm
\begin{center}
{\Large\bf Weak-field approximation of effective gravitational theory with local
 Galilean invariance}\footnote{{\tt rcuzin@phys.ualberta.ca,}{\tt pedro.pompeia@gmail.com,}{\tt montigny@phys.ualberta.ca,}
 {\tt khanna@phys.ualberta.ca} }
\vskip 1cm
{\bf
R. R. Cuzinatto$^{a}$, P. J. Pompeia$^{a,b,c}$, M. de Montigny$^{a,d}$, F. C. Khanna$^{a,e}$ \\}
\vskip 5pt
{\sl $^a$Theoretical Physics Institute, University of Alberta,\\
Edmonton, Alberta, Canada T6G 2J1\\}
\vskip 2pt
{\sl $^b$Instituto de F\'{\i}sica Te\'{o}rica, Universidade Estadual Paulista, \\
Rua Pamplona 145, 01405-000, S\~{a}o Paulo, SP, Brazil \\}
\vskip 2pt
{\sl $^c$Comando-Geral de Tecnologia Aeroespacial, Instituto de Fomento e
Coordena\c{c}\~{a}o Industrial, \\
Pra\c{c}a Mal. Eduardo Gomes 50, 12228-901, S\~{a}o Jos\'{e} dos Campos, SP, Brazil \\}
\vskip 2pt
{\sl $^d$Campus Saint-Jean, University of Alberta, \\
Edmonton, Alberta, Canada T6C 4G9\\}
\vskip 2pt
{\sl $^e$TRIUMF, 4004, Westbrook Mall,\\
Vancouver, British Columbia, Canada  V6T 2A3\\}
\vskip 2pt
\end{center}
\rm
\begin{abstract}
We examine the weak-field approximation of locally Galilean invariant gravitational theories
 with general covariance in a $(4+1)$-dimensional Galilean framework.  The additional degrees
  of freedom allow us to obtain Poisson, diffusion, and Schr\"odinger equations for the
  fluctuation field.   {An advantage of this approach over the usual
  $(3+1)$-dimensional General Relativity is that it allows us to choose an} ansatz for the
   fluctuation field  {that can accommodate the} field
  equations of the Lagrangian approach to MOdified Newtonian Dynamics (MOND) known
  as AQUAdratic Lagrangian (AQUAL).  We investigate a wave solution for the
  Schr\"odinger equations.
\end{abstract}

\end{titlepage}




\section{Introduction to the Galilean framework}\label{Introduction}

It is well known that the Galilei group describes non-relativistic, or low-energy, phenomena
 \cite{LevyLeblond}. The study of non-relativistic limits of various relativistic phenomena
  is not always a straightforward procedure{; for instance, the straightforward low-velocity approximation of electromagnetism
  may lead to one limit, whereas there exist in fact}
  {\it two limits} of electromagnetism \cite{bellac,santos2004}. Another
 example of such a subtlety is the concept of spin; although historically explained within the context
 of the relativistic Dirac equation, spin can also be understood in terms of the Galilean theory,
 so that it is not a purely relativistic notion \cite{leblond}.

The Galilean symmetry is, in some respects, more intricate than the Poincar\'e symmetry,
 which underlies relativistic physics \cite{LevyLeblond}.
  {For instance,} Galilean systems
  are not usually described in terms of tensors, because of the absence of a metric.
  This problem was circumvented, for flat manifolds, with the formalism of  {\it Galilean
   covariance}, originally examined by various authors \cite{takahashi}-\cite{others}. This
  formalism consists in describing Galilean theories in a tensorial form by means of a light-cone metric
   defined on a
 $(4+1)$-dimensional Minkowski manifold. As mentioned in these
 early papers, the reduction from this extended Minkowski space to a
 $(3+1)$-dimensional spacetime can lead to the Galilean invariant theory as
  well as the Minkowski theory,  thereby providing a unified framework to treat both
   the Galilei and the Lorentz kinematics  {in a simple and elegant manner} \cite{omote}-\cite{others}.

In this paper, we extend the formalism to the Riemannian geometry
 underlying gravitational models, following the lines of
  {Ref.} \cite{Schw}.
 One motivation for this work is the fact that General
Relativity (GR) does not provide satisfactory explanations to some
gravitational phenomena involving non-relativistic objects, e.g.
the dark matter and dark energy problems \cite{Cosmology}. The
presence of dark matter is required in order to explain, for
instance, the rotation curves of spiral galaxies{, and} dark energy
appears in the context of cosmology in order to describe the
accelerated expansion of the universe. Although quite diverse, these
issues present a common feature: they occur in the low-energy
regime. This indicates that local Galilean invariance as the
underlying symmetry in the theory of gravitation could shed new
light on these questions. The MOdified Newtonian
 Dynamics (MOND) approach \cite{Milgrom}-\cite{Bekenstein2006}
 addresses these issues either by modifying the law of inertia or
 by altering the Newtonian theory of gravity phenomenologically. Our
 perspective is aligned with the latter approach{; we will show that it accommodates} the
 AQUAL fundamental equation in a rather natural way,
  {with the extra dimension being related to the mass density.}

The Galilean formalism, described in Refs. \cite{takahashi}-\cite{others},
 consists in building Lorentz-covariant action functionals defined on a
 $(4+1)$-dimensional Minkowski manifold. This manifold is
  described in terms of the  \textit{Galilean five-vector}, $(\bx,x^{4},x^{5})$, which
 transforms under Galilei boosts, with relative velocity $\bv$, as
\begin{equation}
\begin{array}[c]{ccl}
\bx^{\prime} & = & \bx-\bv x^{4},\\
{x}^{\prime4} & = & x^{4},\\
{x}^{\prime5} & = & x^{5}-\bv \cdot\bx+\frac{1}{2}{\bv ^{2}}
x^{4}.\label{strans}
\end{array}
\end{equation}
Thus, general Galilean transformations (including rotations, Galilei boosts
and translations) have a form similar to the usual Poincar\'{e}
transformations,
\begin{equation}
x^{\prime\mu}=\Lambda_{\ \nu}^{\mu}x^{\nu}+a^{\nu},\qquad\qquad
\mu,\nu=1,\dots, 5. \label{poincare}%
\end{equation}
The Galilei  {algebra} of a $(3+1)$-dimensional space-time is an
 11-dimensional  {subalgebra} of the 15-dimensional Lie algebra
isomorphic to the Poincar\'{e} algebra in $(4+1)$-dimensional
spacetime.

The tensor methods utilized in the Galilean covariance formalism are based
 on the scalar product, $A\cdot B=\bA \cdot\bB  -A_{4}B_{5} - A_{5}B_{4}$,
 which is invariant under the transformations in Eq. (\ref{strans}).
 This suggests to employ, on a locally flat manifold,
  the following {\it Galilean metric}:
\begin{equation}
\eta_{\mu\nu}=\left(
\begin{array}
[c]{ccc}%
{\mathbf{1}}_{3\times3} & 0 & 0\\
0 & 0 & -1\\
0 & -1 & 0
\end{array}
\right). \label{galileanmetric}%
\end{equation}
We call {\it Galilean manifold} this 5-dimensional flat manifold which, although
 equivalent to a Minkowski manifold in $(4+1)$ dimensions, reduces to the
  Galilean space-time via appropriate ans\"atze.
The {\it five-momentum},
 $p_{\mu}=\mathrm{i}\partial_{\mu}= \left(  \mathrm{i}\nabla,\mathrm{i}%
\partial_{t},\mathrm{i}\partial_{S}\right)  =\left(  \bp, {\frac{E}{c}} ,
mc \right)$,
where $c$ is a parameter with the dimensions of velocity, and $S$ is the
 extra coordinate, suggests that the additional coordinate, $x^{5}=\frac{S}{c}$,
 may be seen as the conjugate of $mc$, where $m$ is the mass.

In this paper, we utilize the Galilean manifold to examine the weak-field approximation
 and we obtain Galilean differential equations through the procedures that usually describe
 gravitational waves in GR.  This will result in Poisson, diffusion, and Schr\"odinger
  equations for the fluctuation field.  {We shall} see in Section \ref{3Ans}
   {that} it is possible to
  accommodate  MOND \cite{Milgrom}-\cite{Bekenstein2006} within the Galilean gravity.
  Henceforth, we consider a
5-dimensional Riemannian spacetime; that is, a 5-dimensional
differential manifold where an invertible symmetric metric tensor is
defined. Thus we can define quantities usually encountered in
general relativity:  a symmetric connection, given by the
Christoffel symbols;  covariant derivatives;  curvature, or Riemann
tensor; Ricci tensor, and the like. The main advantage of the
 Galilean covariance method is that these quantities have a form
  similar to GR, but lead to
   {Galilei-invariant} expressions through a
 natural definition of the ans\"atze.

In Section \ref{WF}, we consider a Lagrangian which produces
equations that are linear in the second-order derivative of the
metric. We also discuss the weak-field approximation with local
Galilean invariance, and obtain field equations. We reduce the
number of degrees of freedom in the perturbation field by using the
trace-reverse perturbation tensor and gauge-fixing conditions. These
results are employed in Section \ref{3Ans}, where we examine three
ans\"atze: the perturbation field does not depend on $x^{5}$, which
leads to a Poisson equation (Section \ref{sss1}); the complex
ansatz, which provides a Schr\"odinger equation (Section
\ref{sss3}); and the real ansatz, which gives a diffusion equation
(Section \ref{sss2}). The ansatz of Section \ref{sss1} may lead to
the AQUAdratic Lagrangian (AQUAL) field equation for MOND theory,
and the ansätze in Section \ref{sss2} suggest a different version
for the AQUAL field equation. We {also} discuss solutions of the
Schr\"odinger equation in Section \ref{sss3}. We make concluding
remarks in Section \ref{Con}.


\section{Weak-field approximation in Galilean gravity}\label{WF}

Typically, the Lagrangian that leads to
  the field equations for  {the point-dependent metric} $g_{\mu\nu} $ is separated into two parts:
$\mathcal{L}_{0}$, which depends on the metric, and
$\mathcal{L}_{\mathrm{matter}}$, which describes a matter field.
The variational principle leads to
 \cite{weinberg,sabbata}
\[
\frac{\delta\mathcal{L}_{0}}{\delta g_{\mu\nu}}=-\frac{\sqrt{-g}}%
{2}\;T^{\mu\nu},\qquad\qquad T^{\mu\nu}\equiv\frac{2}{\sqrt{-g}}\frac
{\delta\mathcal{L}_{\mathrm{matter}}}{\delta g_{\mu\nu}},
\]
where $g\equiv\det(-g_{\mu\nu})$, and
$T^{\mu\nu}$ is the {\it mass-energy-momentum} tensor.
 {Henceforth, greek indices, $\mu$, $\nu$, etc. run from 1 to 5, whereas
 latin indices, $j$, $k$, etc. denote the usual three spatial components.}

 Firstly, we consider the gravitational Lagrangian to be the
Einstein-Hilbert Lagrangian,
\[
{\mathcal{L}}_{0}=\sqrt{-g}\;R,
\]
where $R$ is the scalar curvature. This Lagrangian leads to Einstein's
 equations \cite{weinberg},
\begin{equation}
G_{\mu\nu}\equiv R_{\mu\nu}-\frac{1}{2}g_{\mu\nu}R=\kappa T_{\mu\nu},
\label{EE}%
\end{equation}
where
\begin{equation}
\kappa=8\pi G,
\label{kappa}
\end{equation}
and $G$ is the gravitational constant.  Note that we work with $c=1$.

The following relations are equivalent to Eq. (\ref{EE}):
\begin{equation}
\begin{array}{l}
R_{\mu\nu}=\kappa S_{\mu\nu},\\
S_{\mu\nu}=T_{\mu\nu}-\frac{1}{3}g_{\mu\nu}T, \label{EE1}
\end{array}\end{equation}
where $T=T_{\ \mu}^{\mu}$. Note that the factor $1/3$, instead of
the usual $1/2$, follows from the trace of Eq. (\ref{EE}) which
leads to $R_{\mu\nu}=\kappa\left (T_{\mu\nu}-g_{\mu\nu}\frac 1{D-2}
T\right )$, with $D=5$ instead of $D=4$.

The {\it weak-field}, or {\it linearized-gravity}, approximation is described by the
 {\it perturbation}, or {\it fluctuation}, tensor field
$h_{\mu\nu}$, defined by
\begin{equation}
g_{\mu\nu}=\eta_{\mu\nu}+h_{\mu\nu}, \label{weakfieldg}%
\end{equation}
where $|h_{\mu\nu}|<< 1$. Thus $h_{\mu\nu}$ is the first-order correction to the Galilean metric,
 $\eta_{\mu\nu}$ in Eq. (\ref{galileanmetric}).  The curved spacetime described by
  $g_{\mu\nu}$ is a perturbation of the flat $(4+1)$-dimensional Minkowski manifold.

Consider a finite, global (i.e. $x$-independent) Galilean
transformation,  in the form of Eq. (\ref{poincare}),
\[
x^{\mu }\rightarrow {x}^{\prime\mu }=\Lambda _{\ \nu }^{\mu }\;x^{\nu
}+a^{\mu }, \]
where
$\Lambda _{\rho }^{~\mu }\Lambda _{\sigma }^{~\nu }\eta _{\mu \nu}=\eta _{\rho \sigma }$,
and $\eta_{\mu\nu}$ is the flat-space metric given in Eq. (\ref{galileanmetric}).
The metric transforms as
\begin{align*}
{g'}_{\mu \nu }& =\frac{\partial x^{\rho }}{\partial
{x}^{\prime\mu }} \frac{\partial x^{\sigma }}{\partial
{x}^{\prime\nu }}\ g_{\rho \sigma }=\left( \Lambda ^{-1}\right) _{\ \mu
}^{\rho }\left( \Lambda ^{-1}\right)
_{\ \nu }^{\sigma }g_{\rho \sigma } \\
& =\Lambda _{\mu }^{~\rho }\Lambda _{\nu }^{~\sigma }\left( \eta
_{\rho \sigma }+h_{\rho \sigma }\right) =\eta _{\mu \nu }+\Lambda
_{\mu }^{~\rho }\Lambda _{\nu }^{~\sigma }h_{\rho \sigma }.
\end{align*}
We have used the property $(\Lambda^{-1})^\rho_{\ \
\nu}=\Lambda_\nu^{\ \ \rho}$.

Since we have, in the new frame,
\(
{g'}_{\mu \nu }\left( {x'}\right) =\eta _{\mu \nu}+{h'}_{\mu \nu }\left({x'}\right),
\)
we observe that
 {$h_{\mu\nu}$ transforms like a tensor under Galilean
transformations,}
\[
{h'}_{\mu \nu }=\Lambda _{\mu }^{~\rho }\Lambda _{\nu
}^{~\sigma }h_{\rho \sigma }.
\]

From Eq. (\ref{EE}), we find
\begin{align}
G_{\mu\nu}
& = \frac 12\left[2\;\pd_\rho\pd_{(\mu}h_{\nu)}^{\ \ \rho}
-\pd_\mu\pd_\nu h-\Box h_{\mu\nu}-\eta_{\mu\nu}
\left (\pd_\rho\pd_\sigma h^{\rho\sigma}-\Box h\right )\right]\nonumber\\
&  =\kappa T_{\mu\nu},  \label{Gmunu}%
\end{align}
 where the indices between parentheses are symmetrized; that is,
 \(A_{(\mu\nu)}=\frac 12(A_{\mu\nu}+A_{\nu\mu})\).

We find, from Eq. (\ref{EE1}),  {the equivalent result:}
\begin{equation}
\kappa S_{\mu\nu}=
 \frac 12\left[2\;\pd_\rho\pd_{(\mu}h_{\nu)}^{\ \ \rho}
-\pd_\mu\pd_\nu h-\Box h_{\mu\nu}\right].\label{Smunu}%
\end{equation}
Since $R$ is of first order in $h_{\mu\nu}$, $T_{\mu\nu}$ (or $S_{\mu\nu}$)
must also be of first order. This explains why we use $\eta_{\mu\nu}$,
 instead of $g_{\mu\nu}$, in Eq. (\ref{Gmunu})
   {(or Eq. (\ref{Smunu}))}. In this approximation, the interpretation
of the components of the mass-energy-momentum tensor is manifest. As
 explained in Ref. \cite{santos2004}, the components
$T_{ij}$ are the density of momentum flux of the matter field; $T_{5j}$ is the
field momentum density; $T_{4j}$ is the density of energy flux; $-T_{45}$ the
energy density, and $-T_{55}$ is the density of mass of the field. The component $T_{44}$ does
 not have a clear physical interpretation and it seems to be simply a remnant of the extended
 manifold.

In order to obtain a formal solution of the field equations, let us work with
 the {\it harmonic gauge} condition (see Section 7.4 of Ref. \cite{weinberg}):
\[
g^{\mu\nu}\Gamma_{\mu\nu}^{\rho}=\frac{1}{2}\eta^{\rho\sigma}\left(
2\partial_{\nu}h_{~\sigma}^{\nu}-\partial_{\sigma}h\right) =0.
\]
If we substitute this into the linearized field equation, Eq.
(\ref{Smunu}), and then apply a Fourier transform, we obtain
\begin{equation}
\kappa S_{\mu\nu} (p)= \frac 12 \; p_\rho p^\rho \;  h_{\mu\nu} (p). \label{h(p)}
\end{equation}

By solving Eq. (\ref{h(p)}) for $h_{\mu \nu}(p)$, and performing an
inverse Fourier transform, we obtain
\begin{equation}
h_{\sigma\tau}\left(  x\right)  =2\kappa \int d^{5}x^{\prime}G\left(  x-x^{\prime
}\right)  S_{\sigma\tau}\left(  x^{\prime}\right)  , \label{h(S)}%
\end{equation}
where%
\[
G\left(  x-x^{\prime}\right)  =\int d^{5}p\;e^{\ri p_{\rho}\left(  x-x^{\prime
}\right)  ^{\rho}}\frac{1}{p^{2}}=\int d^{5}p\; e^{\ri p_{\rho}\left(  x-x^{\prime
}\right)  ^{\rho}}\frac{1}{\left(  \bp^{2}-2p_{4}p_{5}\right)  }.
\]
Note that this propagator was used in a different context in Ref.
\cite{Santos2005}. By integrating this expression for the Green's
function, we obtain the field components as integrals of the
energy-momentum tensor, Eq. (\ref{h(S)}).

For later convenience, let us define the {\it trace-reverse perturbation} tensor:
\begin{equation}
\hat{h}_{\mu\nu}\equiv h_{\mu\nu}-\frac{1}{2}\eta_{\mu\nu}h,
\label{TraceReverse}
\end{equation}
so that the  {terms involving the} trace of $\hat{h}_{\mu\nu}$
 {do not appear} in the field
equation. Then we can write Eq. (\ref{Gmunu}), i.e. the field
equations without gauge fixing, as follows:
\begin{equation}
\kappa T_{\mu\nu}=\pd_\rho\pd_{(\mu}\hth_{\nu )}^{\ \ \rho}-\frac 12\eta_{\mu\nu}\;
 \pd_\rho\pd_\sigma\hth^\rho-\frac 12\Box\hth_{\mu\nu}.
\label{kTmunu}
\end{equation}

Furthermore, with the Lorenz gauge  {condition},
 $\partial_{\rho}\hat{h}_{\nu}^{\ \ \rho}=0$,
 we obtain the general field equations
\begin{equation}
\kappa T_{\mu\nu}=-\frac{1}{2}\left[  \nabla^{2}-2\partial_{4}\partial
_{5}\right]  \hat{h}_{\mu\nu}. \label{WE}%
\end{equation}
This equation takes various forms, depending on different hypotheses concerning
the functional form of the field $\hat{h}_{\mu\nu}$.  Specific ans\"atze are discussed
 in Section \ref{3Ans}.

 {(About the gauge fixing, let us comment} that, on the $(4+1)$-dimensional
manifold, the rank-2 symmetric tensor $\hat{h}_{\mu\nu}$ comprises
 15 independent components. The Lorenz
condition, $\partial_{\rho}\hat{h}_{\nu}^{\ \rho}=0$, introduces
five more constraints, and the choice of coordinate system
 implies five additional restrictions. Thus
the linearized tensor has only five independent components:
  $\hth_{1i}$, $\hth_{23}$, $\hth_{33}$.  This will be transparent in Eq.
   (\ref{h TT}) of Section \ref{sss3},
   when we discuss the solutions of the Schr\"odinger equation. { In discussions of gravitational
    waves in five-dimensional GR, one obtains the same number of independent components for
     $h_{\mu\nu}$ (for instance, see Eq. (6) in Ref. \cite{Gre93}). Let us observe that the
    harmonic gauge condition and the Lorentz gauge condition are equivalent; subsequently,
     we choose to use one or the other at convenience.)}



\section{Three ans\"atze for the {perturbation field}}\label{3Ans}

Hereafter we examine three specifc ans\"atze   {of} $\hat{h}_{\mu\nu}$ as a function of $x^5$,
and we
 {examine} Eq. (\ref{WE}),  in the presence of matter-energy fields.


\subsection{Poisson equation}\label{sss1}

If $\hat{h}_{\mu\nu}$ does {\it not depend} on $x^5$, so that
\begin{equation}
\hth_{\mu\nu}=\fhh_{\mu\nu}(\bx,x^4),
\label{ansatzPoisson}
\end{equation}
we see that
Eq. (\ref{WE}) reduces to the Poisson equation:
\begin{equation}
\nabla^{2}\hat{h}_{\mu\nu}\left(  \mathbf{x},x^{4}\right)  =-2\kappa T_{\mu\nu}.
\label{PoisonEq}%
\end{equation}
In vacuum, $T_{\mu\nu}=0$, it simply becomes the Laplace equation.
Notice that, from the point of view of the Galilean approach,
this ansatz means that we are considering $h_{\mu\nu}$ as being a
massless field.

We already mentioned in Section \ref{WF} that $-T_{55}$ is the mass density,
\begin{equation}
-T_{55}=\rho.
\label{T55Rho}
\end{equation}
Then, with Eq. (\ref{T55Rho}) and
\begin{equation}
\hth_{55}=4\phi_N, \label{hath55}
\end{equation}
we find that Eq. (\ref{PoisonEq}) becomes  the Newton's
gravitational law:
 \[
 \nabla^2\phi_N=4\pi G\rho,
 \]
where $\phi_N$ is the Newtonian potential, and $\kappa$ is defined
in Eq. (\ref{kappa}).

Therefore, if we keep Eq. (\ref{T55Rho}), but, instead of
 Eq. (\ref{hath55}), we consider $\hth_{55}$ given by
\begin{equation}
\nabla\hth_{55} = 4 \, \mu\left(x\right)\nabla\phi,\quad
x\equiv|\nabla\phi|/a_{0}\label{h55Aqual}
\end{equation}
then Eq. (\ref{PoisonEq}) gives the Poisson equation utilized in the
AQUAL formalism \cite{AQUAL,Bekenstein2006}:
\begin{equation}
\nabla\cdot\left[\mu\left(x\right)\nabla\phi\right] = 4\pi
G\rho.\label{AQUAL}
\end{equation}

The Milgrom's transition function $\mu$ depends only on the first derivative of the field
 $\phi$, which is the gravitational potential that drives the motion; that is, ${\bf a}=-\nabla\phi$.
  According to MOND, bodies subjected to gravitational force that move with an acceleration smaller than
   $a_0$ should have a dynamic behaviour different from Newtonian mechanics.  AQUAL reproduces
   exactly the MOND formula, $\mu(x){\bf a}=-\nabla\phi_N$, where $\phi_N$ is the Newtonian
   gravitational potential.
  In MOND models, $\mu$
  can, for instance, consist of the ``standard function'', $\mu(x)=x/\sqrt{1+x^2}$, with
   $x= |{\bf a}|/a_0$, which exhibits the additional fundamental constant of
   the theory: the modulus of the acceleration, $a_0\sim 10^{-10}$ m/s$^2$.
 For the standard $\mu$-function given above in the
\textit{large}-$\frac{|\nabla\phi|}{a_{0}}$ limit, we expand $\nabla
h_{55}$ as follows:
\[
\begin{array}{rcl}
\nabla h_{55} & \simeq & \frac{4\nabla\phi}{\sqrt{1+\left(a_{0}/|\nabla\phi|\right)^{2}}}\\
 & = & 4\nabla\phi\;\left[1-\frac{1}{2}\left(\frac{a_{0}}{|\nabla\phi|}\right)^{2}+\frac{3}{8}
 \left(\frac{a_{0}}{|\nabla\phi|}\right)^{4}-\frac{5}{16}\left(\frac{a_{0}}{|\nabla\phi|}\right)^{6}+\cdots\right],\end{array}
 \]
which approaches the Newtonian limit when
$a_{0}<<|\nabla\phi|$. In this limit we can approximate $h_{55}$ by
\[
h_{55}=4\mu\left(x\right)\phi,
\]
since $\nabla
\mu(x)=\frac{d\mu}{dx}\frac{1}{a_{0}\left|\nabla\phi\right|}\left(\nabla\phi\cdot\nabla\right)\nabla\phi,$
and $ \lim_{x\rightarrow\infty}\frac{d\mu}{dx} =
\lim_{x\rightarrow\infty} \frac{1}{\sqrt{(1+x^{2})^3}} = 0$, so that
$\nabla h_{55}=4 \, \mu(x)\, \nabla \phi +4 \phi \, \nabla \mu(x)
\simeq 4 \, \mu(x)\, \nabla\phi$. This identification cannot be done
in the \textit{small}-$\frac{|\nabla\phi|}{a_{0}}$ limit, where we
would obtain
\[
\begin{array}{rcl}
\nabla h_{55} & = & \frac{4\nabla\phi\;|\nabla\phi|/a_{0}}{\sqrt{1+\left(|\nabla\phi|/a_{0}\right)^{2}}}\\
 & = & 4\nabla\phi\;\left[\left(\frac{|\nabla\phi|}{a_{0}}\right)-\frac{1}{2}\left(\frac{|\nabla\phi|}
 {a_{0}}\right)^{3}+\frac{3}{8}\left(\frac{|\nabla\phi|}{a_{0}}\right)^{5}-\frac{5}{16}
 \left(\frac{|\nabla\phi|}{a_{0}}\right)^{7}+\cdots\right],\end{array}
 \]
which approaches the Newtonian limit when
$|\nabla\phi|<<a_0$. In this case $h_{55}$ cannot be integrated and
in order to obtain consistency with the AQUAL theory we should
choose
\[
\nabla\hth_{55}\simeq\;4\mu\left(x\right)\nabla\phi+\nabla\times\mathbf{f},
\]
where $\mathbf{f}$ is a function to be
determined. This is in
accordance to what is expected from the AQUAL version of MOND where
the theory is determined up to a curl of an arbitrary vector field
\cite{Bekenstein2006}.

The same can be done for the ``simple function'', $\mu(x)=x/(1+x)$.
In the \textit{small}-$\frac{|\nabla\phi|}{a_{0}}$ limit, we write
$\nabla h_{55}$ as
\[
\begin{array}{rcl}
\nabla h_{55} & = & \frac{4\nabla\phi\;|\nabla\phi|/a_{0}}{1+|\nabla\phi|/a_{0}}\\
 & = & 4\nabla\phi\;\left[\left(\frac{|\nabla\phi|}{a_{0}}\right)
 -\left(\frac{|\nabla\phi|}{a_{0}}\right)^{2}+\left(\frac{|\nabla\phi|}{a_{0}}\right)^{3}-\left(\frac{|\nabla\phi|}{a_{0}}\right)^{4}+\cdots\right],
\end{array}
\]
which approaches zero in the limit $|\nabla\phi|<<a_{0}$. In this
case, the $h_{55} = \mu(x)\, \phi$ is also valid in first order of
approximation. The simple $\mu$-function, in the
\textit{large}-$\frac{|\nabla\phi|}{a_{0}}$ limit, is obtained by
expanding $\nabla h_{55}$ as
\[
\begin{array}{rcl}
\nabla h_{55} & = & \frac{4\nabla\phi}{{1+a_{0}/|\nabla\phi|}}\\
 & = & 4\nabla\phi\;\left[1-\frac{a_{0}}{|\nabla\phi|}+\left(\frac{a_{0}}{|\nabla\phi|}\right)^{2}
 -\left(\frac{a_{0}}{|\nabla\phi|}\right)^{3}+\cdots\right],
\end{array}
\]
which approaches the Newtonian limit when $a_{0}<<|\nabla\phi|$.
Here a curl of a vector field also should be added in order
to integrate $h_{55}$.


\subsection{Schr\"odinger equation {and its wave solutions}}\label{sss3}

We now turn to the case where the gravitational perturbation
is a massive field. Since $p_{5}$ is an invariant of the Galilean
algebra \cite{santos2004,Santos2005}, the natural ansatz is
given by
\begin{equation}
\hat{h}_{\mu\nu}=e^{-\ri
mx^{5}}\mathfrak{\hat{h}}_{\mu\nu}\left(\mathbf{x},x^{4}\right).
\label{ComplexAnsatz}
\end{equation}
A similar situation occurs in the Galilean analyzes of the Proca
field \cite{santos2004}. In fact, if we follow our recent paper
\cite{Fuad}, and consider both the positive and negative mass
contributions to Eq. (\ref{ComplexAnsatz}), it amounts to adding the
complex conjugate to Eq. (\ref{ComplexAnsatz}), so that
$\hat{h}_{\mu\nu}$ is real. This is analogous to the treatment of
plane waves discussed in Section 10.2 of \cite{weinberg}. The
physical field
$\mathfrak{\hat{h}}_{\mu\nu}\left(\mathbf{x},x^{4}\right)$ should be
real as well. With $\hat{h}_{\mu\nu}$ defined as above, we see that
Eq. (\ref{WE}) becomes
\begin{equation} -2\kappa
T_{\mu\nu}=\nabla^{2}\hat{h}_{\mu\nu}+2m\ri\partial_{4}\hat{h}_{\mu\nu}.\label{schr}\end{equation}
This is an inhomogeneous Schr\"odinger equation for each tensor
component.

Next, we find a solution for the `free-particle' situation, or,
equivalently, the vacuum case, $T_{\mu\nu}=0$.
 This leads to the Schr\"odinger equation, \[
\ri\partial_{4}\hat{h}_{\mu\nu}=-\frac{1}{2m}\nabla^{2}\hat{h}_{\mu\nu}.\]
 If we write the complex ansatz of Eq. (\ref{ComplexAnsatz}) for
$\hat{h}_{\mu\nu}$ in the form, \[
\hat{h}_{\mu\nu}=e^{-\ri mx^{5}}T\left(x^{4}\right)X_{\mu\nu}\left(\mathbf{x}\right),\]
 {then we attain separation of variables in the Schr\"odinger equation.}

If we denote the constant of separation by $E$, the equation for
$T\left(x^{4}\right)$ is \[
\partial_{4}T\left(x^{4}\right)+\ri E\; T\left(x^{4}\right)=0.\]
 Its solution is\begin{equation}
T\left(x^{4}\right)=e^{-\ri Ex^{4}}.\label{T}\end{equation}
 The equation for $X_{\mu\nu}\left(\mathbf{x}\right)$ is \begin{equation}
\nabla^{2}X_{\mu\nu}\left(\mathbf{x}\right)+2mE\; X_{\mu\nu}\left(\mathbf{x}\right)=0.\label{diff eq X}\end{equation}
 The Fourier transform of this equation is \[
\int d^{3}pe^{\ri\mathbf{p\cdot x}}\left[-p^{2}+2mE\right]X_{\mu\nu}\left(\mathbf{p}\right)=0,\]
 which, based on the independence of the plane waves for each value
of $p$, gives the dispersion relation for a non-trivial solution:\begin{equation}
E=\frac{p^{2}}{2m}\;.\label{disp rel}\end{equation}
 This is the non-relativistic energy of a free-particle, as expected,
since we have chosen $T_{\mu \nu} = 0$.

With the dispersion relation, Eq. (\ref{disp rel}), the solution
of Eq. (\ref{diff eq X}) is given by an amplitude $A_{\mu\nu}$ times
the plane wave:\[
X_{\mu\nu}\left(\mathbf{x}\right)=A_{\mu\nu}\; e^{i\mathbf{p\cdot x}}~.\]
 The equation for $X_{\mu\nu}\left(\mathbf{x}\right)$ is solved in
{the} particular gauge given in Section \ref{WF}; that is, $\hth_{i4}=\hth_{44}=\hth_{45}=\hth=0$.
This leads to \[
X_{4i}=X_{44}=X_{45}=0,\]
 and \[
\begin{array}{c}
\nabla^{2}X_{ij}\left(\mathbf{x}\right)+2mE~X_{ij}\left(\mathbf{x}\right)=0,\\
\nabla^{2}X_{i5}\left(\mathbf{x}\right)+2mE~X_{i5}\left(\mathbf{x}\right)=0,\\
\nabla^{2}X_{55}\left(\mathbf{x}\right)+2mE~X_{55}\left(\mathbf{x}\right)=0.\end{array}\]
 The first equation above can be solved in the transverse-traceless
ansatz, usual in the $(3+1)$-dimensional spacetime.

The Fourier transform of the Lorenz-like gauge condition leads to\[
\begin{array}{c}
\ri p_{i}\hth^{i4}-\ri E\hth^{44}=0,\\
\ri p_{i}\hth^{ij}-\ri E\hth^{4j}=0.\end{array}\]
 Then, we have \[
\begin{array}{c}
\hth^{44}=\frac{p_{i}\hth^{i4}}{E},\\
\hth^{4j}=\frac{p_{i}\hth^{ij}}{E}.\end{array}\]
 By substituting the second equation into the first one, we obtain
$\hth^{44}$ in terms of $\hth^{ij}$:\[
\hth^{44}=\frac{p_{i}p_{j}\hth^{ji}}{E^{2}}.\]

Consider a single plane wave; that is, consider a direction of propagation\[
\mathbf{n}=\frac{\mathbf{p}}{\left\vert \mathbf{p}\right\vert }~.\]
 Furthermore take the direction of propagation along the $z$-axis;
that is, \[
p_{i}=p~\delta_{i3}~.\]
 With this restriction, some simplifications follow: \[
\hth^{44}=\frac{p^{2}}{E^{2}}\hth^{33},\]
 and\[
\hth^{4j}=\frac{p}{E}\hth^{3j}~.\]
 Together with the condition, \begin{equation}
\hth=\hth_{11}+\hth_{22}+\hth_{33}=0,\label{traceless}\end{equation}
 this gives \[
\left(\hth_{\mu\nu}^{TT}\right)=\left(\begin{array}{ccccc}
\hth_{11} & \hth_{12} & \hth_{13} & 0 & \frac{p}{E}\hth_{13}\\
\hth_{12} & -\left(\hth_{11}+\hth_{33}\right) & \hth_{23} & 0 & \frac{p}{E}\hth_{23}\\
\hth_{13} & \hth_{23} & \hth_{33} & 0 & \frac{p}{E}\hth_{33}\\
0 & 0 & 0 & 0 & 0\\
\frac{p}{E}\hth_{13} & \frac{p}{E}\hth_{23} & \frac{p}{E}\hth_{33} &
0 & \frac{p^{2}}{E^{2}}\hth_{33}\end{array}\right).\] Note that Eq.
(\ref{traceless}) is also used in the standard relativistic
treatment and it is called ``traceless condition'' in that
relativistic context. The label $TT$, attached to $\hth_{\mu\nu}$,
indicates that we have chosen the {Galilean analogue of the}
transverse-traceless gauge. In terms of the amplitudes, we obtain
\begin{equation}
\hth_{\mu\nu}^{TT}\left(x^{5},x^{4},z\right)=\left(\begin{array}{ccccc}
A_{11} & A_{12} & A_{13} & 0 & \frac{p}{E}A_{13}\\
A_{12} & -\left(A_{11}+A_{33}\right) & A_{23} & 0 & \frac{p}{E}A_{23}\\
A_{13} & A_{23} & A_{33} & 0 & \frac{p}{E}A_{33}\\
0 & 0 & 0 & 0 & 0\\
\frac{p}{E}A_{13} & \frac{p}{E}A_{23} & \frac{p}{E}A_{33} & 0 &
\frac{p^{2}}{E^{2}}A_{33}\end{array}\right)e^{-\ri\left(mx^{5}+Ex^{4}-pz\right)}~.\label{h
TT}
\end{equation}
This is a wave-type solution for the
Schr\"odinger-type equation in the transverse-traceless gauge.

This suggest the following identifications:\[
\omega=E,\qquad\mathrm{{and}}\quad k=\frac{2\pi}{\lambda}=p,\]
 with phase velocity \[
v=\frac{\omega}{k}=\frac{E}{p}.\]
 This result is also obtained in the usual quantum mechanics, described
by the Schr\"odinger equation.

The analogue of Eq. (\ref{h TT}), in the usual {four-dimensional
GR}, exhibits only two degrees of freedom. We can reduce {our five
degrees of freedom} to only two independent entries by setting\begin{equation}
A_{13}=A_{23}=A_{33}=0,\label{A reduction}\end{equation}
 in which case {only $A_{11}$ and $A_{12}$ will remain in the matrix
$\hth_{\mu\nu}^{TT}$.}

The meaning of the extra components in the 5-dimensional spacetime
is still to be clarified. We remark that the additional degrees of
freedom are associated to the massive character of the weak field
solution, manifested by the dependence on the $x^{5}$-coordinate.


\subsection{Diffusion equation}

\label{sss2}

Now we analyze an ansatz analogous to the previous one, but
avoiding the complexification. The gravitational field is still
supposed to be massive, since the dependence with the
fifth-coordinate is kept. If $\hat{h}_{\mu\nu}$ is chosen as
follows,
\begin{equation}
\hat{h}_{\mu\nu}=e^{-mx^{5}}\mathfrak{\hat{h}}_{\mu\nu}\left(\mathbf{x},x^{4}\right),\label{RealAnsatz}\end{equation}
 then Eq. (\ref{WE}) becomes an inhomogeneous diffusion equation,
\begin{equation}
-2\kappa T_{\mu\nu}=\nabla^{2}\hat{h}_{\mu\nu}+2m\partial_{4}\hat{h}_{\mu\nu},\label{dif}\end{equation}
 with $-\frac{1}{2m}$ being the diffusion coefficient. Note that
Eq. (\ref{h55Aqual}), substituted into Eq. (\ref{dif}) produces
an equation similar to Eq. (\ref{AQUAL}) but with an additional time-derivative
of $\hth_{55}$, which is expressed in terms of $\phi$ by integrating
Eq. (\ref{h55Aqual}).

Notice that Eq. (\ref{dif}) is obtained when a gauge is
fixed. If we suppose that the gauge is not fixed and that the field
is static, i.e.
$\mathfrak{\hat{h}}_{\mu\nu}\left(\mathbf{x},x^{4}\right)=\mathfrak{\hat{h}}_{\mu\nu}\left(\mathbf{x}\right)$,
then, from Eqs. (\ref{kTmunu}) and (\ref{RealAnsatz}), we find
\begin{align*}
\kappa T_{44} & =-\frac{1}{2}\nabla^{2}\hth_{44},\\
\kappa T_{55} & =-\frac{1}{2}\left(2m\partial_{i}\hth_{5}^{\ i}+\nabla^{2}\hth_{55}\right),\\
\kappa T_{45} & =\frac{1}{2}\left(m\partial_{i}\hth_{4}^{\
i}+\nabla^{2}\hth_{45}\right).
\end{align*}
The equations in which $\partial_{i}\hth_{4}^{\ i}$ and
$\partial_{i}\hth_{5}^{\ i}$ are non-zero constants reproduce the
Poisson equation with extra terms analogous to the cosmological
constant. This particular choice leads to a \textit{Newton-Hooke
force}; that is, a Newtonian potential plus a (linear) Hooke-like
contribution \cite{Aldrovandi1999}, showing that the
gravitational constant should be related to the mass of the
gravitational field.

Note that it is not possible to obtain such an equation in \textit{four}-dimensional
GR. In principle, it could be done in \textit{five}-dimensional GR,
but this would be far more artificial than the Galilean context utilized
here. Further comments are in Section \ref{Con}.


\section{Concluding remarks}\label{Con}

In this paper, we have retraced the procedures normally leading to the
 investigation of gravitational waves  {in GR}, but we have replaced the Lorentz
 metric with the Galilean metric, Eq. (\ref{galileanmetric}), so that the
 gravitational theory considered is locally Galilean invariant.  The reason
 for doing this is that the Galilei group underlies the appropriate kinematics
 for low-energy, or non-relativistic, phenomena;  therefore, the Galilean symmetry
  must be taken into account when non-relativistic objects are causing
  the gravitational perturbations.

In this paper, we examined the weak-field approximation with local Galilean
 invariance with a $(4+1)$-dimensional Minkowski manifold which leads to
 Galilei-invariant equations, when the physical quantities are properly introduced.
 While there exist 5-dimensional models of GR which have no connection with
 Galilean symmetry, there are a few reasons to consider Galilean gravity.
 For instance, the definition given in Eq. (\ref{h55Aqual}), even though
  permitted in  { five-dimensional} GR, is better justified
  in the Galilean context because Eq. (\ref{T55Rho})
 associates the mass density to the component $T_{55}$ of the energy-momentum
  tensor. Another feature that distinguishes 5-dimensional GR and the Galilean
  formalism employed here is that the ans\"atze, given in Eqs. (\ref{ansatzPoisson}),
  (\ref{RealAnsatz}) and (\ref{ComplexAnsatz}), appear in a natural way within
  the Galilean framework  {\cite{Santos2005}}, whereas no such motivation occurs in GR .

The three ans\"atze considered in Section 3 lead to a Poisson
equation, a diffusion equation, and a Schr\"odinger equation.  The
Poisson equation allows us to introduce the AQUAL field equations
which describe the MOND theory.  This definition and the real anzatz
lead us to suggest a modified version of the AQUAL field equations.
Whether this equation will be relevant in cosmology remains to be
explored.

\section*{Acknowledgments}

We acknowledge partial support by the Natural Sciences and
Engineering Research Council (NSERC) of Canada. RRC and PJP
acknowledge financial support from CNPq (Brazil) and thank the
Physics department, University of Alberta, for providing the
facilities. RRC is grateful to Prof. V P Frolov for the kind
hospitality extended to him. The authors would like to thank
the referee for useful comments.


\begin{thebibliography}{99}

\bibitem{LevyLeblond} L\'evy-Leblond J M 1971  in {\it Group Theory and Applications}, vol. 2
 Ed. E.M. Loebl (New York: Academic Press) p. 221

 \bibitem {bellac} Le Bellac M, L\'evy-Leblond J M 1973 {\it Nuov. Cim.
B} \textbf{14} {217}

\bibitem{santos2004} Santos E S, de Montigny M, Khanna F C, and Santana A E 2004
  {\it J. Phys. A: Math. Gen.} {\bf 37} 9771

\bibitem {leblond} L\'evy-Leblond J M 1967 \textrm{Comm. Math. Phys.}
\textbf{6} {286}

\bibitem {takahashi} Takahashi Y 1988 {\it Fortschr. Phys.} \textbf{36} 63

Takahashi Y 1988 {\it Fortschr. Phys.} \textbf{36} 83

\bibitem {omote} Omote M, Kamefuchi S, Takahashi T, and Ohnuki Y 1989
{\it Fortschr. Phys.} \textbf{37} {933}

\bibitem {others} Soper D E 1976 {\it Classical Field Theory} (Wiley and Sons:
New York) Sect. 7.3

Pinski G 1968 {\it J. Math. Phys} \textbf{9} {1927}

Duval C, Burdet G, K\"unzle H P, and Perrin M 1985 {\it Phys. Rev. D}
\textbf{31} 1841

Duval C, Gibbons G W, and Horv\'athy P 1991 {\it Phys. Rev. D} \textbf{43} 3907

K\"unzle H P and Duval C 1994 in `Semantical Aspects of Spacetime Theories'
(U. Majer and H.J. Schmidt, Eds.) p. 113, BI-Wissenschaftsverlag, Mannheim

Kapu\'scik E 1986 {\it Acta Phys. Pol. B} \textbf{17} 569

\bibitem{Schw} Cuzinatto R R, Pompeia P J, de Montigny M, Khanna F C 2009
 Schwarzschild-type solution in an effective gravitational theory with local Galilean invariance.
  Available at: {\tt{Arxiv: 0903.2488}}

\bibitem{Cosmology} Weinberg S 2008 {\it Cosmology} Oxford, New York

\bibitem{Milgrom}  Milgrom M 1983 {\it Astrophys. J.} \textbf{270} 365

Milgrom M 1983 {\it Astrophys. J.} \textbf{270} 371

Milgrom M 1983 {\it Astrophys. J.} \textbf{270} 384

\bibitem{BekensteinSanders}  Bekenstein J D and Sanders R H 2005 in {\it Mass Profiles and Shapes of
 Cosmological Structures} G. Mamon, F. Combes, C. Deffayet, B. Fort (Eds.), EAS Publications Series.
  Available at: ArXiv: {\tt  astro-ph/0509519}

\bibitem{AQUAL} Bekenstein J D 2004 {\it Phys. Rev. D} {\bf 70} 083509

\bibitem{Bekenstein2006} Bekenstein J 2006 {\it Contemp. Phys.} {\bf 47} 387

\bibitem {weinberg} Weinberg S 1972 {\it Gravitation and Cosmology: Principles and
Applications of the General Theory of Relativity} John Wiley and Sons, New York

\bibitem {sabbata} De Sabbata V and Gasperini M 1985 {\it Introduction to
Gravitation} World Scientific Publishing Co Pte Ltd, Singapore

\bibitem{Santos2005}  {Santos E S, de Montigny M and Khanna F C 2005
{\it Ann. Phys. (NY)} {\bf 320} 21}

 {Abreu L M, de Montigny M, Khanna F C and Santana A E 2003 {\it Ann. Phys. (NY)}
{\bf 308} 244}

\bibitem{Fuad} de Montigny M, Khanna F C and Saradzhev 2008 {\it Ann. Phys. (NY)}
 {\bf 323} 1191

\bibitem{Gre93}  {Gregory R and Laflamme R 1993 {\it Phys. Rev. Lett. } {\bf 70} 2837}

\bibitem{Aldrovandi1999} Aldrovandi R, Barbosa A L, Crispino L C B and
 Pereira J G 1999  {\it Class. Quant. Grav.} {\bf 16}  495




\end{thebibliography}
\end{document}